\documentstyle[aps,pra,psfig,epsf,epsfig,twocolumn]{revtex}

\begin{document}
\draft

\title{Defect band-gap structures for triggering single-photon
emission}

\author{Ho Trung Dung\cite{byline},
Ludwig Kn\"{o}ll, and Dirk-Gunnar Welsch}
\address{
Theoretisch-Physikalisches Institut,
Friedrich-Schiller-Universit\"{a}t Jena,
Max-Wien-Platz 1, 07743 Jena, Germany}

\date{Oct. 14, 2002}
\maketitle

\begin{abstract}
A 3D analysis of the spontaneous decay of a single dipole
embedded in a planar multilayer structure is given, with
special emphasis on Kerr-tunable photonic band-gap materials
for single-photon emission on demand. It is shown that
the change in the density of states near a defect resonance
is much more pronounced than that one near the band edges.
In particular, operation near the band edge as suggested
from a 1D analysis is little suited for controlling
the photon emission.
\end{abstract}


\narrowtext

Quantum states of light hold promise for applications in transmission,
storage, and processing of information in new and powerful ways.
Among them, the single-photon states are of particular interest.
They are crucial in enabling secure transmission
of information without risk of eavesdropping \cite{Gisin02},
and could be useful in all-optical quantum information processing
devices \cite{Knill01}. One of the most natural ways to generate
single-photon states is through controlled radiation of a single
emitter, which, after delivering a photon, is necessarily in the
ground state and will not produce another one before being reexcited.
Various schemes have been reported, such as single-atom passage
through high-$Q$ cavities in the strong-coupling regime
\cite{Maitre97,Varcoe00,Brattke01,Kuhn02},
regulated injection of electron-hole pairs into mesoscopic
quantum wells \cite{Kim99}, and
fluorescence light from individual nitrogen-vacancy color centers
in diamond \cite{Kurtsiefer00,Brouri00,Beveratos02}, from molecules
\cite{DeMartini96,Kitson98,Brunel99,Lounis00,Fleury00,Treussart01},
and from semiconductor quantum dots
\cite{Michler00,Zwiller01,Santori01,Moreau01,Yuan02,Gerard02,Pelton02}.

In practice, single-photon sources that operate in a single-mode
regime and ensure a sufficiently high collection efficiency of the
emitted photon are desired to be used.
A promising route to
achieve these goals is to place the dipole emitters within distributed
Bragg reflector Fabry-Perot \cite{DeMartini96,Kitson98} or pillar
\cite{Moreau01,Gerard02,Pelton02} microcavities.
Due to a strong Purcell selective enhancement of spontaneous
emission, photons are predominantly emitted into the
cavity mode, with a high directionality in the far-field zone.
To truly produce photons one by one, the excitation duration
is commonly required to be short compared to the emitter lifetime.
Note that more than one photon triggered in one shot can, e.g.,
render quantum cryptography insecure under certain types of
eavedropping attacks \cite{Gisin02}.

Recently, a novel `photon-gun' scheme for generating single
photons on demand has been suggested which
is based on the possibility of controlling
spontaneous emission in a photonic band-gap material
\cite{Scheel02}. An initially excited dipole is embedded within a
band-gap structure made of Kerr-nonlinear material such that the
transition frequency of the dipole is inside the band gap,
in fact near its edge, where the spontaneous decay is strongly
inhibited. By applying an external pump field, thanks to Kerr
nonlinearity, the refractive index of the structure
is changed and, as a consequence, the band gap
is shifted in such a way that the dipole transition frequency
falls just outside the gap, into a region of high density
of states. The dipole is then forced to decay rapidly by emitting
a photon. In particular, the scheme offers the advantage of
temporally separating the excitation and the emission process,
thereby putting less constraint on the duration of the excitation.

However, the underlying theory in
\cite{Scheel02} is strictly one-dimensional, thus completely
disregarding the dipole emission into oblique directions.
As we shall show below, this leads to a drastic over-estimation
of the density-of-state switch at the band edge which can be
realized by employing the Kerr effect. Three-dimensional
description of the spontaneous decay of an excited dipole in
planar band-gap structures has been given in
\cite{Kamli00,Li01,Danz02}, with special
emphasis on the limiting case of
an infinitely large number of slabs \cite{Kamli00},
the interference between two decay channels \cite{Li01},
and the dependence of the decay rate
on the position of the dipole \cite{Danz02}.
In all these studies as well as in \cite{Scheel02}
material dispersion and absorption are ignored.

From a 3D theory, we show that there is little chance
of realizing the photon-gun as proposed in \cite{Scheel02}.
We therefore introduce a defect into the band-gap structure
-- a periodic double-layer of quarter-wave plates -- by increasing
the thickness of the layer containing the emitter dipole
from a quarter to a half wavelength. The structure is
now essentially a Fabry-Perot microcavity whose boundaries
are formed by distributed Bragg reflectors, with the defect
resonance inside the band gap being the fundamental resonance
of the cavity. Comparing the operation of the device near the
band edge and near the defect resonance, we show that only in the
latter case a controlled photon emission may be realized.
Finally, we briefly discuss the effect of material absorption
and give an outlook for further improvements.

\begin{figure}[htb]
\noindent
\begin{center}
\epsfig{figure=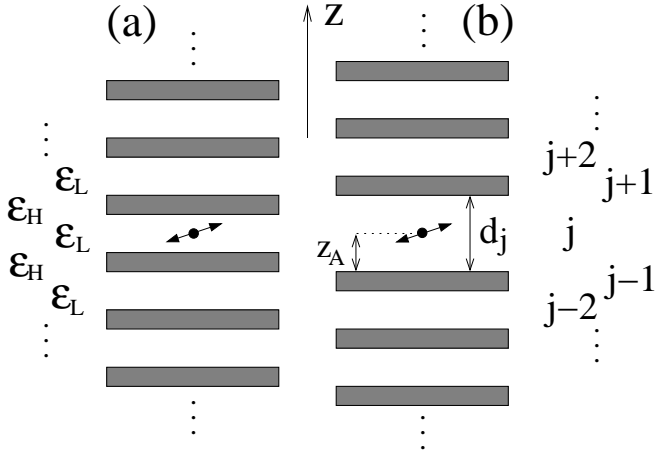,width=\linewidth}
\end{center}
\caption{
A dipole embedded in a band-gap structure (a) with no defect 
and (b) with a defect in the form of a half-wavelength thick layer.
}
\label{geo}
\end{figure}

The system under study consists of a dipole emitter,
which can be an atom, ion, molecule, semiconductor quantum dot,
or nanocrystal, placed in a
photonic band-gap structure consisting of quarter-wave plates
of infinite lateral extension and
periodically interchanging low and high complex permittivities
$\varepsilon_{\rm L}(\omega)$ and $\varepsilon_{\rm H}(\omega)$
(Fig.~\ref{geo}).
The dipole excitation can be realized by using stimulated Raman
adiabatic passage \cite{Bergmann98} with the frequencies of both
branches of the Raman transition lying outside the band gap, or,
in the case of a molecular emitter, by using nonresonant optical
pumping of a vibronic state within the manifold of
vibrational states in the upper electronic
state followed by a fast relaxation to the emission-ready state.

The spontaneous decay rate of a  dipole
(position ${\bf r}_{\rm A}$, transition frequency
$\omega_{\rm A}$, transition dipole moment ${\bf d}_{\rm A}$)
that is surrounded by arbitrary, dispersing and absorbing bodies
can be determined according to the formula
\cite{Agarwal75,Scheel99}
\begin{equation}
\label{e1}
\Gamma = {2\omega_{\rm A}^2\over \hbar\varepsilon_0c^2}\,
     {\bf d}_{\rm A} \,{\rm Im}\,\bbox{G}({\bf r}_{\rm A},{\bf r}_{\rm A},
     \omega_{\rm A})\, {\bf d}_{\rm A},
\end{equation}
where $\bbox{G}({\bf r},{\bf r}',\omega)$ is the
(classical) Green tensor of the medium-assisted Maxwell-field.
Note that $\omega_{\rm A}$ already includes the medium-induced
level shift. The (equal-position) Green tensor can be
decomposed into two parts,
\begin{eqnarray}
\label{e2}
      \bbox{G}({\bf r}_{\rm A},{\bf r}_{\rm A},\omega) =
      \bbox{G}^{\rm bulk}({\bf r}_{\rm A},{\bf r}_{\rm A},\omega) +
      \bbox{G}^{\rm refl}({\bf r}_{\rm A},{\bf r}_{\rm A},\omega),
\end{eqnarray}
where
$\bbox{G}^{\rm bulk}({\bf r}_{\rm A},{\bf r}_{\rm A},\omega)$
is the Green tensor for bulk material and
$\bbox{G}^{\rm refl}({\bf r}_{\rm A},{\bf r}_{\rm A},\omega)$
is the reflection part that insures the correct
boundary conditions at the surfaces of discontinuity.

Let $z$ be the direction of variation of the permittivity of the
multilayer system and the dipole be located in the $j$th layer
(Fig.~\ref{geo}).
Within the frame of the real-cavity model of the
local-field correction, the
part of the dipole decay rate caused by the bulk
Green tensor reads as \cite{Scheel99}
\begin{eqnarray}
\label{e3}
\lefteqn{
     \Gamma^{\rm bulk} = \Gamma_0
     \left|\frac{3\varepsilon_j}
     {2\varepsilon_j\!+\!1}\right|^2
     \Bigg\{ n_{j}'
}
\nonumber \\[.5ex] &&\hspace{1ex}
     +\,\frac{\varepsilon_{j}''}
     {|\varepsilon_j|^2}
     \biggl[
     \left(\!\frac{c}{\omega_{\rm A} R}\!\right)^3
     \!+\!\frac{28|\varepsilon_j|^2
     \!+\!16\varepsilon_{j}'\!+\!1}
     {5|2\varepsilon_j+1|^2}
     \!\left(\!\frac{c}{\omega_{\rm A} R}\!\right)
\nonumber \\[.5ex] &&\hspace{1ex}
     -\,\frac{2}{|2\varepsilon_j\!+\!1|^2}
     \bigl( 2n_{j}''|\varepsilon_j|^2
     +\,n_{j}''\varepsilon_{j}'
     \!+\!n_{j}' \varepsilon_{j}''\bigr)
     \biggr]\Biggr\}
\nonumber \\[.5ex] &&\hspace{1ex}
     +\,O(\omega_{\rm A} R/c),
\end{eqnarray}
where
$\Gamma_0$
$\!= \omega_{\rm A}^3d_{\rm A}^2 / (3\hbar\pi\varepsilon_0c^3)$
is the well-known decay rate in free space,
$\varepsilon_j$ $\!=$ $\!\varepsilon_j(\omega_{\rm A})$
$\!=\varepsilon_{j}'$
$\!+i\varepsilon_{j}''$,
$n_j$ $\!=\sqrt{\varepsilon_j}$
$\!= n_{j}' + in_{j}''$, and
$R$ is the cavity radius that is assumed to be
negligibly small compared with all
characteristic lengths of the configuration under consideration.
The reflection part of the Green tensor can be given in the
form of \cite{Tomas95}
\begin{eqnarray}
\label{e4}
\lefteqn{
     \bbox{G}^{\rm refl}({\bf r}_{\rm A},{\bf r}_{\rm A},\omega)
}
\nonumber\\&&
     = {i \over 4\pi} \int_0^\infty  {\rm d} k_\|\,  k_\|\,
     \frac{e^{i\beta_j d_j}} {2\beta_j}\,
     \tilde{\bbox{G}}{^{\rm refl}}({\bf r}_{\rm A},
     {\bf r}_{\rm A},\omega,k_\|)
\end{eqnarray}
[$k_j$ $\!=$ $\!\sqrt{\varepsilon_j(\omega)}\,\omega/c$;
$\beta_j$ $\!=$ $\!(k_j^2\!-\!k_\|^2)^{1/2}$], where
the nonvanishing components of $\tilde{\bbox{G}}{^{\rm refl}}$ are
\begin{eqnarray}
\label{e5}
    \tilde{G}^{\rm refl}_{xx} = \tilde{G}^{\rm refl}_{yy} =
    -{\beta_j^2\over k_j^2} \, C^p_- +\,C^s_+ ,
    \quad
    \tilde{G}^{\rm refl}_{zz} =
    2 \, {k_\|^2\over k_j^2} \,C^p_+ \,,
\end{eqnarray}
with
\begin{eqnarray}
\label{e6}
    && C^q_{+(-)} = \Bigl[
    r^q_-e^{i\beta_j(2z_{\rm A}-d_j)} + r^q_+e^{-i\beta_j(2z_{\rm A}-d_j)}
\nonumber\\[.5ex]&&\hspace{20ex}
    +\,(-)\, 2 r^q_+ r^q_- e^{i\beta_j d_j}
    \Bigr] D_q^{-1} ,
\\[.5ex]
\label{e7}
    && D_q = 1-r^q_+ r^q_- e^{2i\beta_j d_j}.
\end{eqnarray}
Here, $p(s)$ refers to TM(TE) polarized waves
\mbox{($q$ $\!=$ $\!p,s$)}, and
$r^q_{+(-)}$ are the total reflection coefficients at the
upper (lower) stack of layers, which
are to be calculated using recurrence formulas \cite{Tomas95}.
In the numerical calculation, material
dispersion and absorption are taken into
account through a permittivity of Drude-Lorentz type
\begin{equation}
\label{e8}
        \epsilon(\omega) = 1 +
        {\omega_{\rm P}^2 \over
        \omega_{\rm T}^2 - \omega^2 - i\omega \gamma}\,,
\end{equation}
where $\omega_{\rm P}$ corresponds to the coupling constant,
and $\omega_{\rm T}$ and $\gamma$ are respectively the medium oscillation
frequency and the linewidth. Note that for small absorption, the pole
singularities in the Sommerfeld integrals appearing in Eq.~(\ref{e4})
remain close to the Re($k_\|$) axis and might cause serious
problems. This can be avoided by deforming the
integration path in the complex plane \cite{Paulus00}.

\begin{figure}[htb]
\noindent
\begin{center}
\epsfig{figure=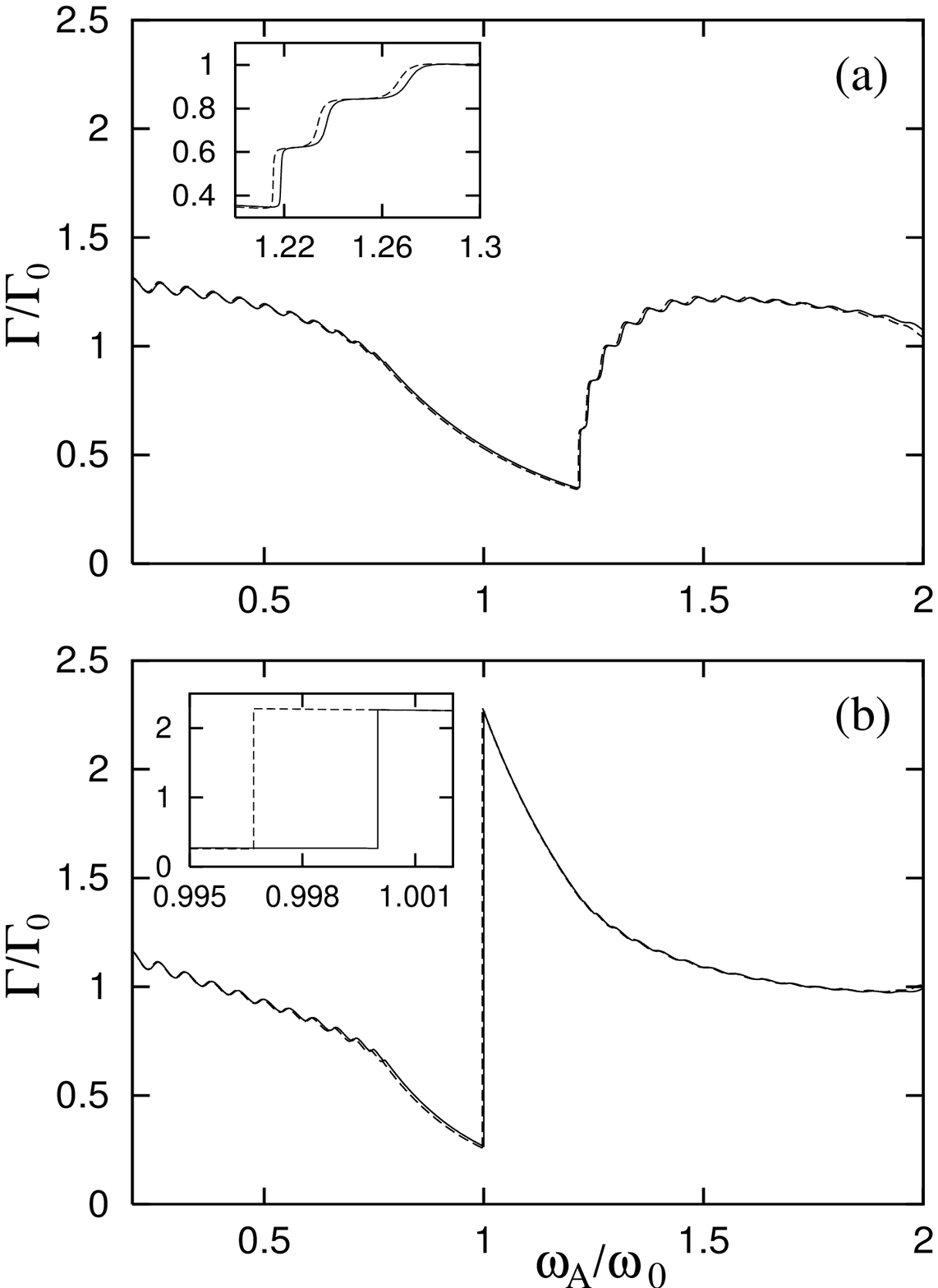,width=1.\linewidth}
\end{center}
\caption{
(a) The decay rate of an $x$-oriented dipole
located at the center of
the middle layer of a 30-period structure
is shown as a function of the transition frequency
for \mbox{$\varepsilon_j$ $\!=\varepsilon_{\rm L}$ $\!=1$}
and $\varepsilon_{\rm H}(\omega)$ from Eq.~(\ref{e8}) with
\mbox{$\omega_{\rm T}$ $\!=$ $\!20\,\omega_0$},
\mbox{$\gamma$ $\!=$ $\!10^{-7}\,\omega_0$}, and
\mbox{$\omega_{\rm P}$ $\!=$ $\!1.7299\,\omega_{\rm T}$}
[\mbox{$\varepsilon_{\rm H}(\omega_0)$ $\!\simeq$
$\!4+i\,7.5 \times 10^{-10}$}; solid line] and
\mbox{$\omega_{\rm P}$ $\!=$ $\!1.7529\,\omega_{\rm T}$}
[\mbox{$\varepsilon_{\rm H}(\omega_0)$ $\!\simeq$
$\!4.0804+i\,7.7 \times 10^{-10}$}; dashed line].
(b) Normalized decay rate when a defect is introduced in the
form of a half-wavelength thick middle layer. The
other parameters are the same as in Fig.~\ref{def}(a)
}
\label{def}
\end{figure}

An example of the frequency response of the
rate of spontaneous decay of a dipole in a band-gap structure
according to Fig.~\ref{geo}(a) is shown in Fig.~\ref{def}(a).
It is seen that near the band edge the decay rate
can be regarded as being a multiple-step function of
frequency rather than a single-step function
as in a 1D theory \cite{Scheel02}.
In particular, a change of the real part of the
permittivity $\varepsilon_{\rm H}(\omega_0)$ from
\mbox{$\varepsilon_{\rm H}'(\omega_0)$ $\!=$ $\!4$}
to \mbox{$\varepsilon_{\rm H}'(\omega_0)$ $\!=$ $\!4.0804$}
($\omega_0$, mid-gap frequency)
is seen to change the decay rate,
for \mbox{$\omega_{\rm A}$ $\!=$ $1.217\,\omega_0$}, from
\mbox{$\Gamma$ $\!=$ $\!0.35\,\Gamma_0$} to
\mbox{$\Gamma$ $\!=$ $\!0.6\,\Gamma_0$}, which
is far below the change from \mbox{$\Gamma$ $\!\simeq$ $\!0$}
to \mbox{$\Gamma$ $\!=$ $\!50\,\Gamma_0$}, as it is predicted
from the 1D theory in \cite{Scheel02}.
In this context it should be pointed out that
in the 1D theory the change of the spontaneous decay
near the band edge significantly increases with the
number of layers, which is in
strong
contrast to the 3D theory
where the available change effectively decreases.

To improve the performance of the device, we suggest to introduce
a defect in the form of a half-wave plate and to operate it near
the resulting defect resonance rather than near the band edge.
The frequency response of the decay rate observed in that case
is illustrated in Fig.~\ref{def}(b).
For the same change of the permittivity as in Fig.~\ref{def}(a),
the decay rate now changes from
\mbox{$\Gamma$ $\!=$ $\!0.25\,\Gamma_0$} to
\mbox{$\Gamma$ $\!=$ $\!2.26\,\Gamma_0$}, that is
a change of one order of magnitude becomes feasible.
Moreover, the change is much more abrupt than in Fig.~\ref{def}(a),
which is of course an added bonus.

\begin{figure}[htb]
\noindent
\begin{center}
\epsfig{figure=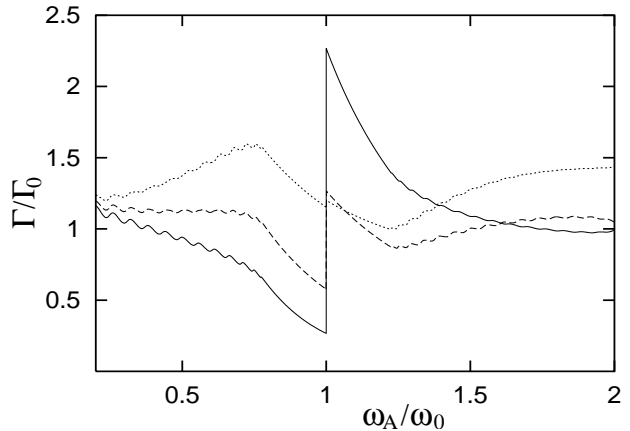,width=1.\linewidth}
\end{center}
\caption{
The transition frequency dependence
of the dipole decay rate in Fig.~\ref{def}(b)
is shown for different dipole positions:
\mbox{$z_{\rm A}$ $\!=0.5d_j$} (solid line),
\mbox{$z_{\rm A}$ $\!=0.2d_j$} (dashed line), and
\mbox{$z_{\rm A}$ $\!=0.05d_j$} (dotted line).
The permittivity $\varepsilon_{\rm H}(\omega)$ is
given according to Eq~(\ref{e8}) with
\mbox{$\omega_{\rm T}$ $\!=$ $\!20\,\omega_0$},
\mbox{$\gamma$ $\!=$ $\!10^{-7}\omega_0$}, and
\mbox{$\omega_{\rm P}$ $\!=$ $\!1.7299\,\omega_{\rm T}$}.
The other parameters are the same as in Fig. \ref{def}(b).
}
\label{pos}
\end{figure}%

{F}igure \ref{pos}
illustrates the influence of the position of the dipole on
the frequency dependence of the decay rate
for a defect-multilayer structure according to Fig.~\ref{geo}(b).
It is seen that when the dipole is located in the middle of the
layer, then the change of the decay rate at the defect resonance is
most abrupt and thus best suited for the device operation.

\begin{figure}[htb]
\noindent
\begin{center}
\epsfig{figure=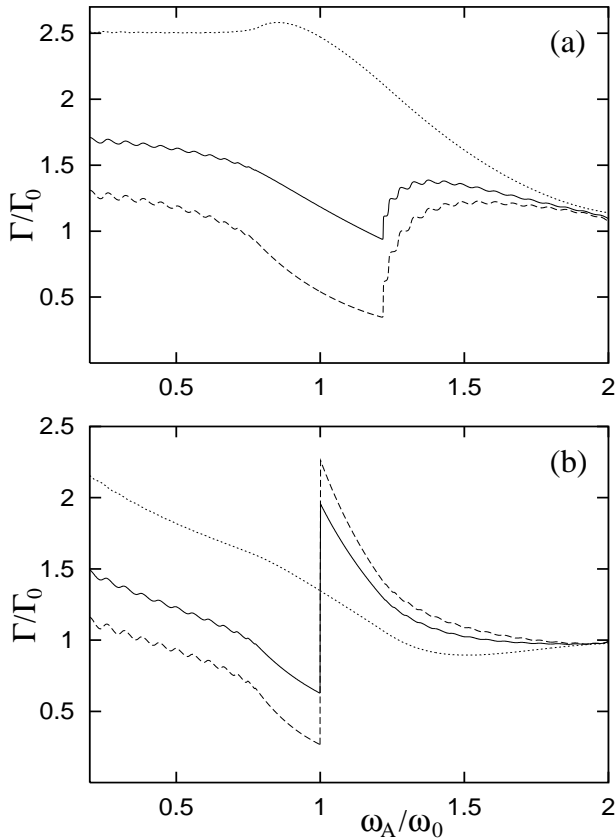,width=1.\linewidth}
\end{center}
\caption{
Decay rate for an $x$-oriented dipole (dashed line),
a $z$-oriented dipole (dotted line), and the averaged decay rate
[\mbox{$\Gamma$ $\!=$ $\frac{1}{3}(2\Gamma_x+\Gamma_z)$};
solid line].
Case (a) refers to the band-gap structure without defect
[Fig.~\ref{geo}(a)], while case (b) refers to the structure
with the defect [Fig.~\ref{geo}(b)].
The dipole is located at the center of the middle layer.
The other parameters are the same as in Fig.~\ref{pos}.
}
\label{orient}
\end{figure}

So far we have assumed that the dipole orientation is
parallel to the layers, e.g., in $x$-direction.
In practice, this may not always be the case.
In Fig.~\ref{orient}, we have plotted the frequency response of
the decay rate for an $x$- and a $z$-oriented dipole
and the decay rate averaged over all possible
dipole orientations.
Let us assume that $\varepsilon_j(\omega_{\rm A})$ can be
regarded as being real.
Decomposing the integral in Eq. (\ref{e4}) into two parts,
\mbox{$\int_0^{k_j} \ldots$ $\!+\int_{k_j}^\infty\ldots$},
we can distinguish between propagating and evanescent waves
in the $z$-direction. Numerical computations (not shown)
indicate that the abrupt change of the decay rate of an
$x$-oriented dipole near the band edge
[Fig.~\ref{orient}(a) for a structure according to Fig.~\ref{geo}(a)]
and near the defect resonance
[Fig.~\ref{orient}(b) for a structure according to Fig.~\ref{geo}(b)]
is closely related to the case that the decay
is associated with propagating-wave excitation,
in which case the emitted photon
can really escape the structure and subsequently be collected.
On the contrary, a $z$-oriented dipole is stronger coupled
to the evanescent field than to the propagating field,
and thus does not feel the
(qua\-si\mbox{-)f}or\-bidden frequency range (see Fig.~\ref{orient}).
Since this undesired effect is more pronounced
for the perfect band-gap structure
[Fig.~\ref{orient}(a)]
than for the structure with the defect in
[Fig.~\ref{orient}(b)],
a single-photon source
operating near a defect resonance is much more robust against
a randomization of the dipole orientation than the one operating
near the band edge of a defect-free structure.

\begin{figure}[!t!]
\noindent
\begin{center}
\epsfig{figure=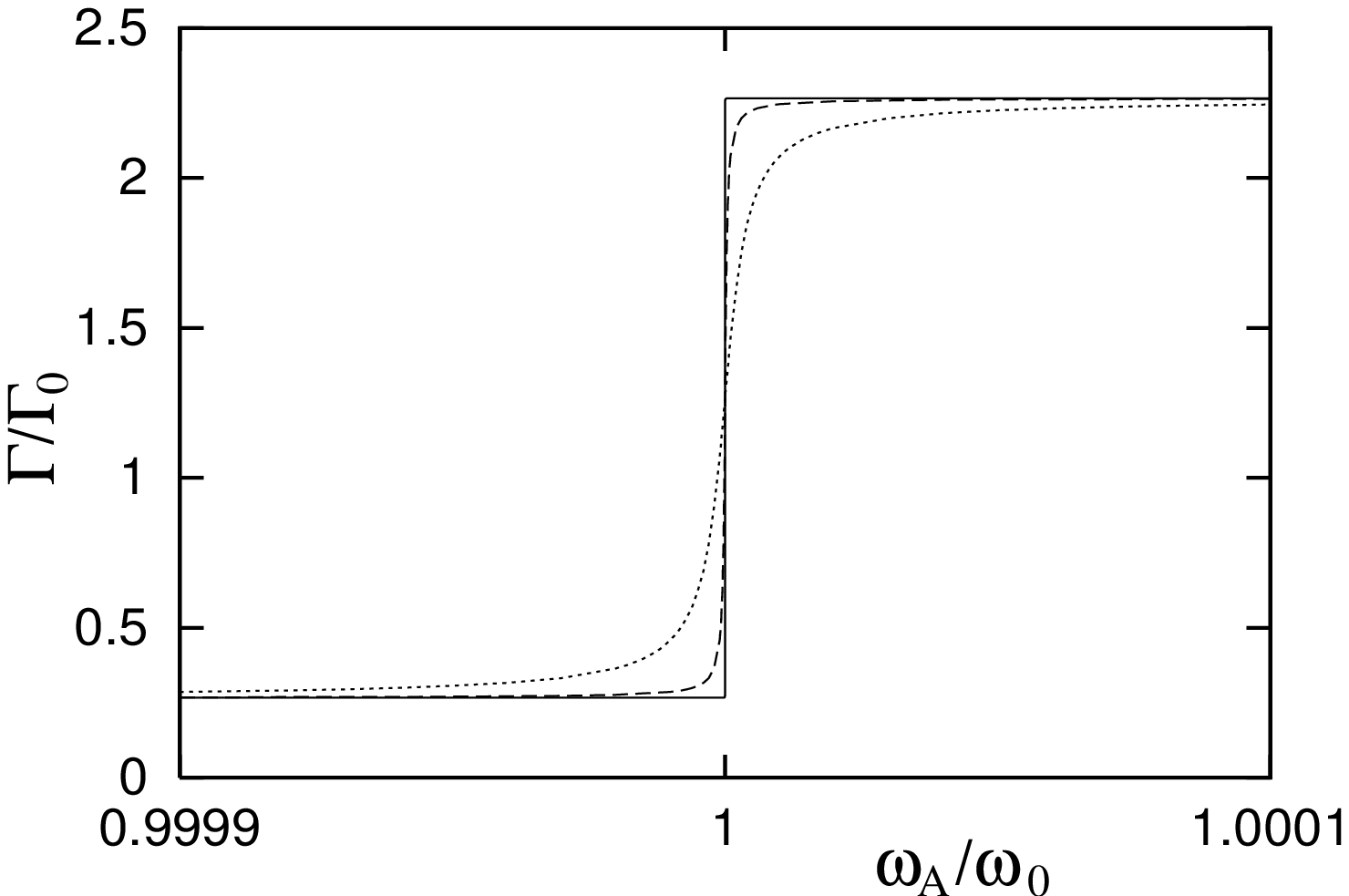,width=1.\linewidth}
\end{center}
\caption{
The effect of material absorption on  the dipole decay rate
in Fig.~\ref{def}(b) is shown for $\varepsilon_{\rm H}(\omega)$
according to Eq.~(\ref{e8}) with
\mbox{$\gamma$ $\!=$ $\!10^{-7}\omega_0$} (solid line),
\mbox{$\gamma$ $\!=$ $\!10^{-3}\omega_0$} (dashed line), and
\mbox{$\gamma$ $\!=$ $\!10^{-2}\omega_0$} (dotted line)
[\mbox{$\omega_{\rm T}$ $\!=$ $\!20\,\omega_0$},
\mbox{$\omega_{\rm P}$ $\!=$ $\!1.7299\,\omega_{\rm T}$}].
The other parameters are the same as in Fig.~\ref{def}(b).
}
\label{absp}
\end{figure}

In Fig.~\ref{absp} the effect of material absorption on the
decay rate near the defect resonance is illustrated.
As expected, absorption tends to smooth the change of the
decay rate at the defect resonance.
For $\gamma$ $\!=$ $\!10^{-7}\,\omega_0$ and
a relative change of $\varepsilon_{\rm H}'(\omega_0)$ of
the order of $10^{-2}$, the band-gap shift is about
$10^{-3}\,\omega_0$ [see the inset of Fig. \ref{def}(b)].
For a more realistic relative
change of $\varepsilon_{\rm H}'(\omega_0)$, say, of the order of
$10^{-3}$ achievable by the optical Kerr effect, numerical
computations (not shown) indicate a smaller band-gap shift of about
$10^{-4}\,\omega_0$. Even in this case, the band-gap shift is still
much larger than the modification caused by (small) material
absorption, as it is seen from Fig. \ref{absp}. Therefore, one
can safely say that reasonably small material absorption does not
significantly affect the device operation.
It should be noted that the spontaneous decay rate given
by (\ref{e1}) is the total rate, which
takes into account radiative decay via both
propagating and evanescent waves and, for absorbing matter,
also nonradiative decay.
For a detailed analysis, a better measure of the efficiency
of the photon emission process
may be the far-zone emission pattern and the (total) amount
of radiative energy sent out.

Although we have given numerical examples only for symmetric
configurations, the theory is applicable for arbitrary
planarly stratified media.
Nonsymmetric structures with a better reflecting wall
on one side, which can improve the light
collection efficiency on the other side of the structure,
are of special interest.
Another possible improvement of the device
is a better photon confinement, e.g., by enclosing the
emitter in micropillars or two-, or three-dimensional band-gap
structures, so that a more drastic switching of the photonic
density of states can be achieved.
Besides the intensity-dependent Kerr nonlinearity,
fast switching of photonic density of states in photonic crystals
can also be achieved by using, e.g., two-photon excitation of
free carriers \cite{Johnson02}.

The authors acknowledge discussions with S. Scheel.
This work was supported by the Deutsche Forschungsgemeinschaft.



\begin{references}
\bibitem[*]{byline} On leave from the Institute of Physics, National Center
for Natural Sciences and Technology, 1 Mac Dinh Chi Street,
District 1, Ho Chi Minh city, Vietnam.

\bibitem{Gisin02}
N. Gisin, G. Ribordy, W. Tittel, and H. Zbinden,
Rev. Mod. Phys. {\bf 74}, 145 (2002).

\bibitem{Knill01}
E. Knill, R. Laflamme, and G. J. Milburn,
Nature (London) {\bf 409}, 46 (2001).

\bibitem{Maitre97}
X. Ma{\^{\i}}tre, E. Hagley, G. Nogues, C. Wunderlich, P. Goy,
M. Brune, J. M. Raimond, and S. Haroche,
Phys. Rev. Lett. {\bf 79}, 769 (1997).

\bibitem{Varcoe00}
B. T. H. Varcoe, S. Brattke, M. Weidinger, and H. Walther,
Nature (London) {\bf 403}, 743 (2000).

\bibitem{Brattke01}
S. Brattke, B. T. H. Varcoe, and H. Walther,
Phys. Rev. Lett. {\bf 86}, 3534 (2001).

\bibitem{Kuhn02}
A. Kuhn, M. Hennrich, and G. Rempe,
Phys. Rev. Lett. {\bf 89}, 067901 (2002).

\bibitem{Kim99}
J. Kim, O. Benson, H. Kan, and Y. Yamamoto,
Nature (London) {\bf 397}, 500 (1999).

\bibitem{Kurtsiefer00}
C. Kurtsiefer, S. Mayer, P. Zarda, and H. Weinfurter,
Phys. Rev. Lett. {\bf 85}, 290 (2000).

\bibitem{Brouri00}
R. Brouri, A. Beveratos, J.-P. Poizat, and P. Grangier,
Opt. Lett. {\bf 25}, 1294 (2000).

\bibitem{Beveratos02}
A. Beveratos, S. K{\"u}hn, R. Brouri, T. Gacoin, J.-P. Poizat,
and P. Grangier,
Eur. Phys. J. D {\bf 18}, 191 (2002).

\bibitem{DeMartini96}
F. De Martini, G. Di Giuseppe, and M. Marrocco,
Phys. Rev. Lett. {\bf 76}, 900 (1996).

\bibitem{Kitson98}
S. C. Kitson, P. Jonsson, J. G. Rarity, and P. R. Tapster,
Phys. Rev. A {\bf 58}, 620 (1998).

\bibitem{Brunel99}
C. Brunel, B. Lounis, P. Tamarat, and M. Orrit,
Phys. Rev. Lett. {\bf 83}, 2722 (1999).

\bibitem{Lounis00}
B. Lounis and W. E. Moerner,
Nature (London) {\bf 407}, 491 (2000).

\bibitem{Fleury00}
L. Fleury, J.-M. Segura, G. Zumofen, B. Hecht, and U. P. Wild,
Phys. Rev. Lett. {\bf 84}, 1148 (2000).

\bibitem{Treussart01}
F. Treussart, A. Clouqueur, C. Grossman, and J.-F. Roch,
Opt. Lett. {\bf 26}, 1504 (2001);
F. Treussart, R. All{\'e}aume, V. Le Floc'h, L. T. Xiao,
J.-M. Courty, and J.-F. Roch,
Phys. Rev. Lett. {\bf 89}, 093601 (2002).

\bibitem{Michler00}
P. Michler, A. Imamo\v{g}lu, M. D. Mason, P. J. Carson, G. F. Strouse,
and S. K. Buratto,
Nature (London) {\bf 406}, 968 (2000);
P. Michler, A. Kiraz, C. Becher, W. V. Schoenfeld, P. M. Petroff,
L. Zhang, E. Hu, and A. Imamo\v{g}lu,
Science {\bf 290}, 2282 (2000).

\bibitem{Zwiller01}
V. Zwiller, H. Blom, P. Jonsson, N. Panev, S. Jeppesen, T. Tsegaye,
E. Goobar, M.-E. Pistol, L. Samuelson, and G. Bj{\"o}rk,
Appl. Phys. Lett. {\bf 78}, 2476 (2001).

\bibitem{Santori01}
C. Santori, M. Pelton, G. Solomon, Y. Dale, and Y. Yamamoto,
Phys. Rev. Lett. {\bf 86}, 1502 (2001).

\bibitem{Moreau01}
E. Moreau, I. Robert, J. M. G{\'e}rard, I. Abram, L. Manin,
and V. Thierry-Mieg,
Appl. Phys. Lett. {\bf 79}, 2865 (2001).

\bibitem{Yuan02}
Z. Yuan, B. E. Kardynal, R. M. Stevenson, A. J. Shields, C. J. Lobo,
K. Cooper, N. S. Beattie, D. A. Ritchie, and M. Pepper,
Science {\bf 295}, 102 (2002).

\bibitem{Gerard02}
J. M. G{\'e}rard, B. Gayral, and E. Moreau,
quant-ph/0207115.

\bibitem{Pelton02}
M. Pelton, C. Santori, J. Vuckovic, B. Zhang, and G. S. Solomon,
J. Plant, Y. Yamamoto, quant-ph/0208054.

\bibitem{Scheel02}
S. Scheel, H. H{\"a}ffner, H. Lee, D. V. Strekalov, P. L. Knight,
and J. P. Dowling, quant-ph/0207075.

\bibitem{Kamli00}
A. Kamli and M. Babiker, Phys. Rev. A {\bf 62}, 043804 (2000).

\bibitem{Li01}
G. X. Li, F. L. Li, and S. Y. Zhu,
Phys. Rev. A {\bf 64}, 013819 (2001).

\bibitem{Danz02}
N. Danz, R. Waldh{\"a}usl, A. Br{\"a}uer, and R. Kowarschik,
J. Opt. Soc. Am. B {\bf 19}, 412 (2002).

\bibitem{Bergmann98}
K. Bergmann, H. Theuer, and B. W. Shore,
Rev. Mod. Phys. {\bf 70}, 1003 (1998).

\bibitem{Agarwal75}
G. S. Agarwal, Phys. Rev. A {\bf 12}, 1475 (1975);
J. M. Wylie and J. E. Sipe, Phys. Rev. A {\bf 30}, 1185 (1984);
Ho Trung Dung, L. Kn\"{o}ll, and D.-G. Welsch,
Phys. Rev. A {\bf 62}, 0538041 (2000).

\bibitem{Scheel99}
S. Scheel, L. Kn\"{o}ll, and D.-G. Welsch,
Phys. Rev. A {\bf 60}, 4094 (1999).

\bibitem{Tomas95}
M. S. Toma\v{s}, Phys. Rev. A {\bf 51}, 2545 (1995);
W. C. Chew, {\it Waves and Fields in Inhomogeneous Media}
(IEEE Press, New York, 1995).

\bibitem{Paulus00}
M. Paulus, P. Gay-Balmaz, and O. J. F. Martin,
Phys. Rev. E {\bf 62}, 5797 (2000).

\bibitem{Johnson02}
P. M. Johnson, A. F. Koenderink, and W. L. Vos,
Phys. Rev. B {\bf 66}, 081102(R) (2002).

\end{references}
\end{document}